\documentclass[twocolumn,showpacs,preprintnumbers,amsmath,amssymb]{revtex4}
\usepackage{graphicx}% Include figure files
\usepackage{dcolumn}% Align table columns on decimal point
\usepackage{bm}% bold math

\begin{document}

\title{Forward Raman compression via photonic band gap in metals or warm dense matter}% \\

\author{S. Son}
\affiliation{18 Caleb Lane, Princeton, NJ 08540}
\author{Sung Joon Moon\footnote{Current Address: 28 Benjamin Rush Ln., Princeton, NJ 08540}}
\affiliation{Program in Applied and Computational Mathematics, Princeton University, Princeton, NJ 08544}
\date{\today}% It is always \today, today,

\begin{abstract}
The group velocity of a light pulse in photonic band gap material
could considerably deviate from the speed of light in vacuum.
Different speeds of a forward stoke and a pump pulse would enable
the Raman compression in metals or the warm dense matter.     
A small window of the parameter regime, where the compression is feasible
via the forward Raman scattering, is identified.   
\end{abstract}


\maketitle

The photonic band gap (PBG) material, consisting of periodic dielectric or
metallo-dielectric structures, has strong interaction with the photons,
leading to rich phenomena. 
It is widely used to manipulate the visible lights for various practical
purposes~\cite{Yablo}.
It is shown that even the soft x-ray could be manipulated using the PBG~\cite{Kue}.
In particular, it is noted that the group velocities of the pulses could be 
considerably different from each other.

Manipulating an x-ray becomes increasingly more important and practical,
as advanced light sources make intense x-rays available for many
applications~\cite{fast4, Free2,Tabak}.    
The backward Raman scattering (BRS), which has been successful for the compression
of the visible light, could make such intense x-rays even shorter and more intense
in dense plasmas~\cite{Fisch3, Fisch, McKinstrie,Ping}. 
However, 
many physical processes to be considered for the x-ray compression in dense plasmas
are different from those for visible lights in ideal plasmas,
including the Fermi degeneracy~\cite{sonprl,sonpla}, 
the electron quantum diffraction~\cite{sonlandau}, 
the electron band gap~\cite{Kue} and 
the plasmon decay~\cite{sonbackward}, to mention a few.
These different processes render some severe difficulties in the visible light
BRS become harmless in the x-ray BRS, or vice versa.
For example, the parasitic forward Raman scattering (FRS)
might not be severe in dense plasmas~\cite{Gibbons,malkin1, Sturm,Sturm2,sonbackward}. 

The goal of this paper is to apply the photonic band gap concept and the aforementioned
peculiar properties of dense plasmas to the forward Raman x-ray compression. 
%Based on the plasmon decay theory recently developed~\cite{sonbackward2} and  
% the different speed of the pulses in the photonic band gap material, 
We propose that the FRS could be used as a compressing mechanism,
as opposed to being an obstacle for the compression. 
The thresholds for the pulse intensity and duration required for the compression are estimated.

We begin by considering a band gap material,
consisting of alternating layers of two different kinds of metals (denoted by metal A and B below).
Let $\delta \omega_{pe}^2 = 4\pi(n_A-n_B)e^2/m_e$ where $n_A$ ($n_B$) is the electron density of 
the layer A (B). 
As discussed in Ref.~\cite{Kue}, the band width of the pump pulse should be 
much smaller than the band gap, in order for a well-defined wave packet 
of the velocity to be considerably different from speed of the light in vacuum.
This condition imposes a constraint on the pulse duration~\cite{Kue} 
\begin{equation}
\tau > 10 \left( \frac{\omega}{\omega_{\mathrm{pe}}} \right)^2 
\frac{1}{\omega} \label{eq:duration} \mathrm{,}
\end{equation}
where $\omega^2_{\mathrm{pe}} = 4 \pi n_B e^2 / m_e$ and $\omega$ is the frequency of the light pulse. 
When the above conditions are satisfied, the FRS can be used as a compressing mechanism,
as the seed and the pump pulses could travel with different velocities. 
If the seed and pump travel with the same velocity, the FRS cannot be used as a compressing channel.

The one-dimensional (1-D) equations describing the three-wave interactions of the Raman
scattering in the PBG material reads~\cite{McKinstrie}:
\begin{eqnarray}
\left( \frac{\partial }{\partial t} + v_1 \frac{\partial}{\partial x} + \nu_1\right)A_1  = -ic_1 A_2 A_3  \nonumber \\
\left( \frac{\partial }{\partial t} + v_2 \frac{\partial}{\partial x} + \nu_2\right)A_2  = -ic_2 A_1 A^*_3  \nonumber \\
\left( \frac{\partial }{\partial t} + v_3 \frac{\partial}{\partial x} + \nu_3\right)A_3  = -ic_3 A_1 A^*_2  
\label{eq:2},
\end{eqnarray}
where $A_1 = eE_1/m\omega_1c$ ($A_2$) is the electron quiver velocity of the pump (seed) 
pulse scaled by the speed of light $c$, $A_3 = \tilde{n_e}/n_e$ is the Langmuir wave intensity;
$\nu_1 $ ($\nu_2$) is the rate of the inverse bremsstrahlung of the pump (seed),
and $\nu_3$ is the rate of the plasmon decay;
$c_1 = \omega_{pe}^2/2\omega_1$, $c_2 = \omega_{pe}^2/2\omega_2$,
$c_3 = (ck_3)^2/2\omega_3$; 
$\omega_{1} (\omega_{2})$ is the frequency of the pump (seed),
and $\omega_3=\omega_{\mathrm{pe}}= (4\pi n_e^2/m_e)^{1/2}$. 
The energy conservation relation in the FRS (BRS)
is given as $\omega_1 = \omega_2 + \omega_{pe}$ and 
the momentum conservation is $ k_3 = k_1 - k_2 $ ($k_1+k_2$). 
The plasmon decay rate $\nu_3(q)$ in metals, 
which is mainly attributed to the Umklapp processes~\cite{Sturm}, is much higher
than the theoretical predictions based on the random phase approximation.
% or the dynamical correlation theory~\cite{DeBois}.  %% 
%The conventional theories~\cite{Landau,Lindhard,Hammett,DeBois}  
%predict the vanishing damping  as $q$ goes to zero, 
% but the experiments suggest that the decay rate is finite~\cite{Gibbons}.  
%This is due to the Umklapp processes such 
%as the inter-band transition~\cite{Hasegawa,Sturm}.
%Without further details, we approximate 
The plasmon decay rate for $q < 0.5 k_F$, where $k_F$ is the Fermi wave vector,
is given as 
\begin{equation}
\nu_3(q) = \eta(q) \omega_{\mathrm{pe}}  \label{eq:pla} \mathrm{,}
\end{equation}
where $\eta(k) = \eta_0 + d\eta/dq^2 (q/k_F)^2$.  
For typical metal, $ 0.02 < \eta_0 < 0.2$ and 
$k_F^2  d\eta/dq^2 \cong a \eta_0$ ($2<a<10$)~\cite{Gibbons}.
More detailed discussion can be found in Ref.~\cite{sonbackward2}.
The inverse bremsstrahlung ($\nu_1$ and $\nu_2$) in metal is~\cite{Shima, son2}
\begin{equation}
\nu_B(\omega, E) = 4 \pi n_i Z_i^2 \left( \frac{e^4}{m_e^2 v_F^3} \right) 
\frac{\omega_{\mathrm{pe}}^2}{\omega^2} \frac{F(\alpha)}{ s^2\kappa^{1/2}}
 \mathrm{,} \label{eq:brem}
\end{equation} 
where it is assumed that $ \kappa =  (\hbar \omega / 2 m_e v_F^2) < 1 $.
$v_F$ is the Fermi energy and $ \alpha^2 = 2e^2E^2/m\hbar \omega^3 $,
where $E$ is the electric field of the pulse. 
%and we divide by factor 2 for the above  formula  because the most of the inverse bremsstrahlung is coming from the denser layer in the PBG material. 
$F(\alpha)$ can be approximated as $F(\alpha) \cong \alpha^2/6 $ when $\alpha < 2$
and $F(\alpha) \cong (2/\pi\alpha)\log(2\alpha)^2$ when $\alpha \ge 2$ \cite{Shima}.

In order for  the pump to overcome the inverse bremsstrahlung,  the minimum pump pulse duration $\tau$ given by Eq.~(\ref{eq:duration}) should be larger than the inverse bremsstrahlung decay time. 
This condition reads 
\begin{equation}
\nu_B(\omega, E)\tau =\frac{ 40 \pi n_i Z_i^2 \left( \frac{e^4}{m_e^2 v_F^3} \right)}{\omega_1} \frac{F(\alpha)}{ s^2\kappa^{1/2}}
\ll 1 \label{eq:m} \mathrm{.}
\end{equation}
which is satisfied for most metal when $\omega_1 /\omega_{\mathrm{pe}} > 5$. 
From Eq.~(\ref{eq:2}), the linear Raman growth condition from the FRS is estimated to be 
$ c_2 c_3 |A_1|^2 >  \nu_2 \nu_3$, where $c_1$, $c_2$ and $c_3$ are assumed to be smaller
by a factor of 2 compared to what is given in Eq.~(\ref{eq:2})
(provided that the Raman interaction mainly arises from the dense layers).   
Then the threshold criteria for the FRS instability is given as 
\begin{equation}
A_{1F}^2 = \frac{ 8.91\times 10^{16}Z\sec}{ \omega_{pe}} \eta(q) s_2^{-3/2} \log(2\alpha_2)^2 (\frac{\omega_{pe}}{\omega_2}) \label{eq:3} \mathrm{.}
\end{equation} 
where it is assumed that $\alpha > 2$.
The condition, $ A_1 > A_{1F}$, is necessary for the compression. 
%Unless the seed pulse intensity is such that $ A_1 > A_{1F}$,  there will be no compression from the FSR. 
For example, consider Aluminum (AL) with $\omega_1 /\omega_{\mathrm{pe}} = 10$ ($Z=3$, $\eta \cong 0.03$, and  $\omega_{\mathrm{pe}} = 2.32 \times 10^{16}\  \sec^{-1} $). 
%In this case,  we can simplify Eq.~(\ref{eq:3})  as 
% $ A_{1F}^2 = 2.303 \eta s_2^{-3/2} \log(2\alpha_2)^2$.  
For $s_2 = 50$, which corresponds to the electron quiver energy (due to the seed pulse) of $ 500 \ \mathrm{eV}$ and  the laser intensity of $
3.04 \times 10^{19} \ \mathrm{W}/\mathrm{cm}^2$, the instability threshold of the pump field is  $0.8 \times 10^{19} \  \mathrm{W}/\mathrm{cm}^2$.

There could be a parasitic FRS from the background noise plasmon. 
As the noise seed is not intense enough ($\alpha < 2 $),
we use an approximate functional form $F(\alpha) \cong \alpha^2/6 $. 
The growth rate of the noise instability from Eq.~(\ref{eq:2}) is 
\begin{equation} 
\gamma= \frac{1}{16\eta}\frac{\omega_{pe}}{\omega_1}A_1^2 \omega_{pe}\mathrm{.} \label{eq:growth}
\end{equation} 
Assuming  the seed  has the duration of $\tau_p$, 
 the background FRS can have a growth factor at most $\exp(\tau_p \gamma)$. 
For $\omega_{pe}/\omega = 0.1 $ and $\eta= 0.03$, 
this is estimated to be $\exp(0.2 A_1^2 (\omega_{pe}\tau_p))$.
As long as  $\tau_p < 50 / A_1^2$,  the background FRS cannot deplete the pump,
which is the case due to the condition Eq.~(\ref{eq:m}). 
%If the pump intensity is higher than the instability threshold give in Eq.~(\ref{eq:th}), 
%the parasitic FRS could becomes in principle unstable.
% However, the alternating layer will prevent this instability. 
Furthermore, the alternating layer would prevent this instability. 
The background plasmon (the pump) in the FRS has a wave vector of 
$\omega_{\mathrm{pe}}/c$ ($\omega_1 / c $), where $\omega_1 /\omega_{\mathrm{pe}} \geq 5$.  
Let us assume that each layer has an excitation of a plasmon of the form 
 $E_i \exp(ik_3y - i\omega_{pe}t+i\theta_i)$, 
where $\theta_i$ is an independent phase between the layers. 
%the excitation is given as $E_i \exp(ik_3y - i\omega_{pe}t+i\theta_i)$.  
%For these plasmons to  excite  the FRS loop,
The amplitude of $|\Sigma E_i E_j\exp(i(\theta_i-\theta_j))|^2$
is not  proportional to $N^2$ but to  $N$, where  $ N \cong 2 \omega_1 /\omega_{\mathrm{pe}}$.
%  because  $\theta_i$ are indepenent one another. 
This lack of coherence  prevents the background noise FRS. 
%Even if the background parasitic FRS is unstable,
% unless the first stoke grows from the background to a point where it depletes the pump,   
%the effect of the background FRS will be negligible.

\begin{figure}
%\scalebox{1.0}{
%\includegraphics[width=0.895\columnwidth]{pump_forward}}
\scalebox{1.0}{
\includegraphics[width=0.9\columnwidth]{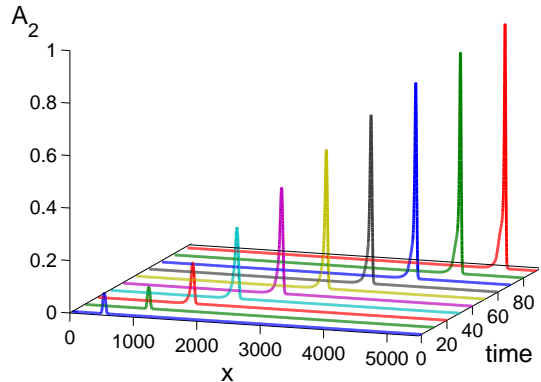}}
\caption{\label{fig:pump} The time evolution of the seed pulse,
where $t$ is normalized by $ \omega_{\mathrm{pe}} = 2.39 \times 10^{16} / \sec $. 
The pump has duration of $4000/\omega_{\mathrm{pe}}$, which corresponds to 173 femto second, 
and the electron quiver energy of $0.8 \ \mathrm{keV}$ ($0.6 \times 10^{20} \ \mathrm{W}/\mathrm{cm}^2$).  
The seed has the duration of 4 femto seconds with the quiver energy of $1.6 \ \mathrm{keV}$.
The PBG material is located in $500< x <5000$, where $x$ is normalized 
by $ c/ \omega_{\mathrm{pe}} =1.30 \times 10^{-6} \mathrm{cm}$.  
%The y-axis is the quiver velocity divided by the velocity of the light.   
 }
\end{figure}

We demonstrate the estimation given by Eqs.~(\ref{eq:m}) and (\ref{eq:3}) 
through 1-D simulation of Eq.~(\ref{eq:2}) applied to AL.  
In the numerical integration of Eq.~(\ref{eq:2}) (Fig.~\ref{fig:pump};
the initial seed and pump pulses propagate to the right), 
where the seed is amplified by a factor of 80 in its intensity.
It is assumed that the pump pulse has the group velocity $c/2$ and 
the seed has the velocity of $c$, which can be achieved by choosing
the pump pulse frequency slightly below the band gap.

In the above simulation, it is assumed that the pump is slower than the seed (slow-pump):
the seed pulse sweeps through the pump pulse and extracts the energy.
It is also possible that the seed would be slower than the pump (slow-seed).
% the pump sweeps through  the seed. 
%In the slow-seed case,  
%it is harder for the seed to satisfy  Eq.(~\ref{eq:m}), 
%but  the growth rate for the parasitic FRS from the second stoke
%  is weaker
% because the seed and the second stoke have different group velocities each other. 
 %In the slow-seed case, the part of the pump that freshly 
%e%ntering the PBG material will give the energy to 
%the seed before its energy is dissipated by the inverse bremsstrahlung. 
For a highly intense pump, the inverse bremsstrahlung is a serious concern,
where the slow-seed would be preferable because the part of the pump newly 
entering the PBG material will give out the energy to the seed before its energy
gets dissipated by the inverse bremsstrahlung. 
For a moderately intense pump, the slow-pump might be better as it is easier
to make the pump slower than the seed in the PBG material.

To summarize, the possibility of using the forward Raman scattering for the x-ray
compression is examined.
Our scheme is based on the fact that, in the PBG material, the forward stoke could
have a different group velocity than the pump pulse. 
% The first  stoke could move faster or slower than the seed pulse extracting energy.  
The background FRS would no longer be a big concern as a consequence of the strong decay of Langmuir waves.   
On the other hand, the strong inverse bremsstrahlung will put a limit on the pump pulse 
duration and increase the threshold intensity of the pump for Raman growth. 
We identify the plausible pulse characteristics as given by
Eqs.~(\ref{eq:duration}), (\ref{eq:m}) and (\ref{eq:3}).

In metal, the backward Raman scattering could be also plausible.
% due to the fact that the FRS is less harmful.  
Even though the excitation of the plasmon in the BRS is
 stronger than that in the FRS by a factor of
 $4(\omega/\omega_{pe})^2  ( \eta(\omega_{\mathrm{pe}}/c)/\eta(2\omega/c))$,  
 the BRS instability from the noise plasmons is  
a severe problem.
Furthermore,
there is a maximum frequency limit ($2 \omega/c \ll k_F$), over which the BRS compression is impossible. 
The FRS does not have such issues and therefore it is advantageous compared to the BRS in a very intense x-ray compression.

\end{document}